# A REAL-TIME DATABASE QOS-AWARE SERVICE SELECTION PROTOCOL FOR MANET

Jihen Drira Rekik[1] Leïla Baccouche[1] and Henda Ben Ghezala[1]

[1] RIADI-GDL laboratory, ENSI, National School of Computer Sciences

Manouba University, 2010 Manouba, Tunisia

## ABSTRACT

*The real-time database service selection depends typically to the system stability in order to handle the time-constrained transactions within their deadline. However, applying the real-time database system in the mobile ad hoc networks requires considering the mobile nodes limited capacities. In this paper, we propose cross-layer service selection which combines performance metrics measured in the real-time database system to those used by the routing protocol in order to make the best selection decision. It ensures both timeliness and energy efficiency by avoiding low-power and busy service provider node. A multicast packet is used in order to reduce the transmission cost and network load when sending the same packet to multiple service providers. In this paper, we evaluate the performance of our proposed protocol. Simulation results, using the Network Simulator NS2, improve that the protocol decreases the deadline miss ratio of packets, increases the service availability and reduces the service response time.*

## 1. INTRODUCTION

With the rapid growth of the real-time constrained information services used in a mobile database environment, there is an increasing demand to support and guarantee the quality of service (QoS) such as completing the transactions within their deadline. The mobile ad hoc network, MANET, is of interest because there is no prior investment for fixed infrastructures, it can be easily deployed in a short time and end users can access and manipulate data anytime and anywhere. Each mobile node acts as a host autonomous end system (requestor or source data node) and as a router (intermediate nodes) for others in the network. The mobile nodes are classified by their capacities into two groups: small mobile host SMH, requestor node, which has a reduced memory, storage, power and computing capabilities and large mobile host LMH, source data node, equipped with more storage, power and communication and computing facilities than the SMH. The databases are stored at source data, LMH, and accessed by requestor nodes, SMH. The key challenging in the mobile networks is providing the time constrained services with node limited resources. For example, the military applications have to respect time constraint in order to update positions of wounded soldiers, get enemy map position or find medical assistance. If the wounded warriors cannot reach a medical assistance center in time, the information might become useless. For another example, the information about source data overload that is caused by important number of user transactions sent to it should also reach a mobile requestor heading towards this direction timely. If the requestor receives such information early enough, it will be able to react accordingly to avoid the overloaded source data node.

Currently, most MANET research has focused on routing and connectivity issues in order to cope with the dynamism of such networks. Just solving the problem of connectivity is not sufficient. These applications often need times to utilize resources or services that are present on other mobile nodes. Often nodes cannot communicate directly with each other. MANET's basic role is to allow mobile users to exchange data and use each other's database services. So there are needs for data communication protocol and database service selection protocol without any central intelligence in the network. Otherwise, the node lifetime is usually affected by the unpredictable user transactions to the source data node (service provider). This constraint may lead to overload and so to deplete the source data node energy which may affect the service availability. With the frequent disconnections due to the energy depletion, the transactions may miss their deadlines whereas there are other same service providers available in the vicinity. Service selection of the nearest database service provider is insufficient. Other criteria for database service selection should be highlighted. It must guarantee a minimum of QoS essential for the execution of transactions with respect to the deadline. However, respecting the deadline cannot be insured nor guaranteed if the energy resource is exhausted, or with a system overloaded by unpredictable transactions from other mobile nodes. There has been a significant research interest towards semantic, context and/ or QoS-aware service selection. We pay particular attention to QoS-aware service selection as it matches with our research interests. We focus especially on cross-layer service selection a special class of efficient service selection approaches for MANET. The service selection algorithm combines QoS parameters according to different layers. Cross-layer service selection exploits the capability of integrating service information along with routing information (in the same message), thus the service selection packets at the application layer are avoided and energy is saved. To find a suitable database service in MANET is a challenging problem and is relative to the appropriate QoS metrics choice and their relevance.

In this paper, we aim to apply the cross-layer service selection for selecting an available real-time database system, RTDBS. The workload in these systems is unpredictable. They may become quickly overloaded and unstable leading to the decrease of the required quality of service (QoS) in terms of executing transactions within their deadlines. Therefore, we pay particular attention to integrate the performance metrics measured in RTDBS to those used by routing protocol in order to make the best selection decision, under the deadline constraint. How to guarantee the transaction without exceeding its real-time constraints or wasting resources? What are the appropriate metrics combining timeliness on the one hand and on the other hand choosing the suitable service without depleting resources? We propose Real-Time Database QoS-aware Service Selection protocol (RTDQS). In RTDQS, the database service selection is done based on both intermediate nodes and database service provider QoS metrics. Then, we aim to evaluate cross-layer service selection proposed protocol and its ability to reduce energy consumption, reduce service response time and guarantee timeliness.

The rest of the paper is organized as follows: in the second section, we present the related work to the service selection protocols. Then, we present the QoS guarantees in a RTDBS in the third section. In the next section, we describe the proposed Real-Time Database QoS-aware Service Selection protocol (RTDQS). Detailed analysis of system performance is performed in the sections V and VI.

## 2. RELATED WORK

A number of survey paper regarding service discovery protocols for mobile ad-hoc network has been done to present their merits and drawbacks in [12] and [13]. The service discovery is defined as a process allowing networked entities to advertise their services, to query about services provided by other entities, to select the most appropriately matched services and to invoke the services [13]. Service selection is a basic feature for service discovery approaches. In this paper, we focus especially on the cross-layer service selection protocols. It relies on International Journal of Database Management Systems ( IJDMS ) Vol.3, No.4, November 2011interaction between the network and the application layer information. In network layer based service selection protocol, the routing messages are enhanced with service information. Therefore, with cross-layer service selection protocol, the control overhead is reduced [10]. There has been a significant research interest towards semantic matching or QoS-aware service selection. We pay particular attention to QoS-aware service selection. With QoS awareness, generally, a selection algorithm is based on certain criteria or metrics. These metrics join route specific (e.g., hop-count, bandwidth, delay) and service specific (e.g., service provider load, remaining energy, capacity).

In [6], Jayapal and al. propose an adaptive service selection protocol. The service provider selection is based on both the distance between the service requestor and the service provider and the remaining lifetime of the service provider. In [14], Zhang and al. propose a cross-layer approach to service selection. The service provider is selected according to a cost function which depends to service information such as response time and battery level. In [3], the authors study the impact of two basic service selection strategies. The first one studies the impact on the lifetime of the service providers where a client always selects the nearest one with less hops. The second one studies the impact on the whole network where a client always selects the service provider with a maximum remaining energy. The service provider's remaining energy yields the best performance when is used as a service selection criterion than the shortest path selection criterion.

We note that in the above solutions the service selection is only based on the service provider autonomy such as the service provider's remaining energy. It considers in no way the link state such as path's energy, neither the service provider load. In fact, the link state reflects the intermediate mobile nodes capacities to maintain the data link connection. We may select a stable service provider with high capacities whereas some intermediate mobile nodes, used to reach the service provider, have limited capacities.

## 3. QOS WITH THE REAL-TIME DATABASE SYSTEM

QoS with the RTDBS has a different meaning than the one with MANET. RFC 2386 [4] characterizes the QoS for a network as a set of service requirements to be met by the network while transporting a flow from one source to a destination such as the delay, the energy efficient, the bandwidth and the packet loss. However, with the RTDBS, the QoS aims to guarantee that the transactions are completed within their deadlines and, on the other hand, to provide robustness and the stability of the system against unpredictable workload variations. The workload is unpredictable due to user transactions. The system may become quickly overloaded, leading to the decrease of the RTDBS performance criterion. Several approaches focus on a transaction differentiation according to a deadline criticality [5]. Others one focus on the QoS parameters observed on the RTDBS to provide robustness against unpredictable workload variations [7], [2].

### 3.1. QoS according to the transaction type

Gruenwald and al., [5], presents a Power Efficient Transaction management technique for Real-time mobile Ad-hoc NETwork databases, called PETRANET. It takes into account the energy limitations, the client/server mobility and the real-time constraints imposed by the transactions of a real-time mobile ad-hoc network database. The requestor needs to send the transaction to an appropriate source data, called the server, for processing. For the transaction with soft deadline, the requestor selects the source data with the highest residual energy level. However, the firm transactions are sent to the nearest server. The goal of PETRANET is to reduce the number of transaction aborts while maintaining a balance of energy consumption distribution among mobile

103

nodes. The first deadline of soft transactions is neglected in favor of balancing the energy consumption. The soft transaction is still executed after the first deadline has expired.

### 3.2. QoS according to the observed parameters

Several approaches using a RTDBS have been proposed to handle real-time constrained transactions. Kang and al. [7] and Amirijoo and al. [2] propose the quality of service parameters in order to quantify system's performance. These parameters can specify the deadline miss ratio or the transient overloads period. To specify the required performance of a real-time database system, they define threshold values for transient-state performance metrics such as overshoot (V) and settling time (T), not to exceed. V is the worst-case system performance in the transient system state and it is given in percentage. T is the time for a transient miss ratio overshoot to decay and reach the steady state performance. It is a measure of how fast the system converges towards the desired performance. After reaching the settling-time T, the real-time database should enter the steady state (as seen in figure 1). It is a measure of system adaptability.

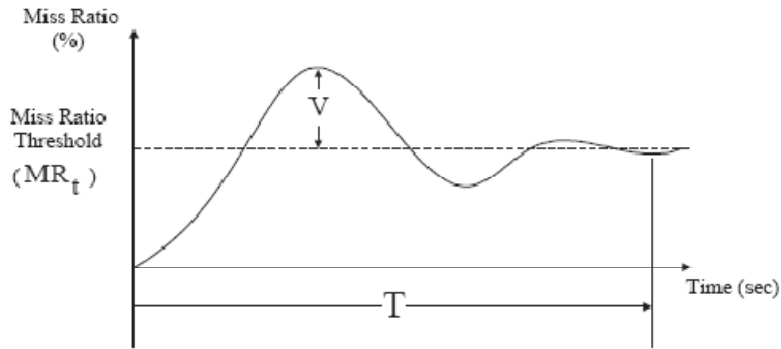

Figure 1. T Transient performance metrics

Kang and al. [7] and Amirijoo and al. [2] adopt both steady-state and transient-state performance metrics. The following table summarize performance metrics and the correspondent reference values as threshold values.

Table 1. QoS specification

| QoS Metrics | | Reference Values |
|---|---|---|
| **User transactions quality metrics** | Deadline Miss Ratio (MR) | ≤ 10% |
| | Average Transactions Error (ATE) | ≤ 20% |
| **Transient-state performance metrics** | Overshoot (V) | ≤ 30% |
| | Settling-Time (T) | ≤ 60 sec |

The transactions miss ratios is defined as follows:

$$MR = 100 \times \frac{\text{Tardy}}{\text{Terminated}} (\%)$$

Where Tardy denotes the number of transactions that have missed their deadline; Terminated is the number of terminated transactions.

The average transaction error defines the preciseness of the results of user transactions (transactions) and defined as:

$$ATE(k) = \frac{\sum_{i \epsilon Terminated(k)} TE_i}{Terminated(k)}$$

Where Terminated(k) denotes the set of terminated transactions; $TE_i$ is the error of the transaction $T_i$. It represents the accuracy of the results provided by the transaction.

## 4. THE PROPOSED CROSS-LAYER SERVICE SELECTION PROTOCOL

None of the above works exploits the information related to the RTDBS status in database service selection in terms of a database system workload and stability. The main problem is to choose the reliable and the efficient RTDBS service provider to handle the real-time transactions (with respect to their deadlines) within MANET constraints. The service selection should also consider the exhaustible mobile nodes resources of the service requestor, service provider and intermediate mobile nodes, too.

We are interested in the problem of service selection when the same service is provided by multiple large mobile hosts as source data node. We pay particular attention to cross-layer service selection. The cross-layer service selection algorithm combines QoS parameters according to application layer and routing layer in order to compute route and to select the appropriate RTDBS service. In routing layer based service selection protocol, the routing messages are enhanced with service information.

As routing protocol, we rely on Energy and Delay aware - DSR, ED-DSR, proposed in our last work [11], in order to take into account the state of the intermediate mobile nodes for each connection transaction. For the proposed real-time database QoS aware service selection protocol, RTDQS, we have made the modifications to the underlying ED-DSR protocol's route discovery and the route selection mechanisms. The RTDQS introduces two extra messages types Service Request, SREQ, and Service Reply, SREP.

In the next subsections, we present the basic routing protocol for route selection. Then, we describe the proposed cross-layer service selection protocol.

### 4.1. An energy and delay - dynamic source routing protocol

The choice of the suitable route in ED-DSR [11] is conditioned by three factors: the residual energy of nodes belonging to the route, the delay requirements of the real-time flow and the load of the intermediate node's queue. The selected route should satisfy delay requirements, preserve the energy consumption and avoid the overloaded nodes. Therefore, each intermediate mobile node stamps its current status in the RREP packet. ED-DSR calculates the cost of each available route according to the following equation:

$$C_{Routing} = \sum_{i=1}^{M} \left( \alpha \times C_{energy}^i + \beta \times C_{queue}^i + \gamma \times C_{delay}^i \right)$$

Where i is the index of the intermediate mobile nodes in the route path, $C_{Routing}$ is the cost of the route and $C_{energy}^i$, $C_{queue}^i$ and $C_{delay}^i$ are the cost functions describing the node i characteristics such as the energy, queue length and delay, respectively. α, β and γ are the costing factors which normalize $C_{energy}^i$, $C_{queue}^i$ and $C_{delay}^i$. $C_{energy}^i$ is a function relative to the distance and the remaining energy of the node i. $C_{queue}^i$ is relative to the queue length along the current route. $C_{delay}^i$ depends to the local processing time of each packet in each queue along the current route. A route is selected based on minimum values of the $C_{Routing}$.

Each intermediate node verifies if the packet can reach the destination before the expiration delay. Otherwise, the node discards the route. The path should respect and guarantee the deadline.

$$D_k > \sum_{i=1}^{N} C_{delay}^i.$$

Where $D_k$ is the worst case execution time for the packet k with N intermediate mobile nodes.

## 4.2. How the RTDQS works in MANET database

The RTDQS introduces two extra messages types the Service Request, SREQ, and the Service Reply, SREP. The format of our service selection protocol packet is slightly different from the original ED-DSR in order to introduce the QoS parameters of RTDBS service in application layer. We modified the SREP packet format and added an extra field in the packet format of ED-DSR to store the service cost function, $C_{QoS}$, relative to the actual status of the service provider node, LMH. The $C_{QoS}$ is described in the next service selection subsection.

| Next Header | | Reserved | Payload Length |
|---|---|---|---|
| Options | Data | $C_{QoS}$ | $C_{Delay}$ |

Figure 2. The SREP packet format of RTDQS including metrics for service selection

The figure 3 represents the flow chart of the proposed real-time database QoS aware service selection protocol RTDQS highlighting our contribution in bold chart.

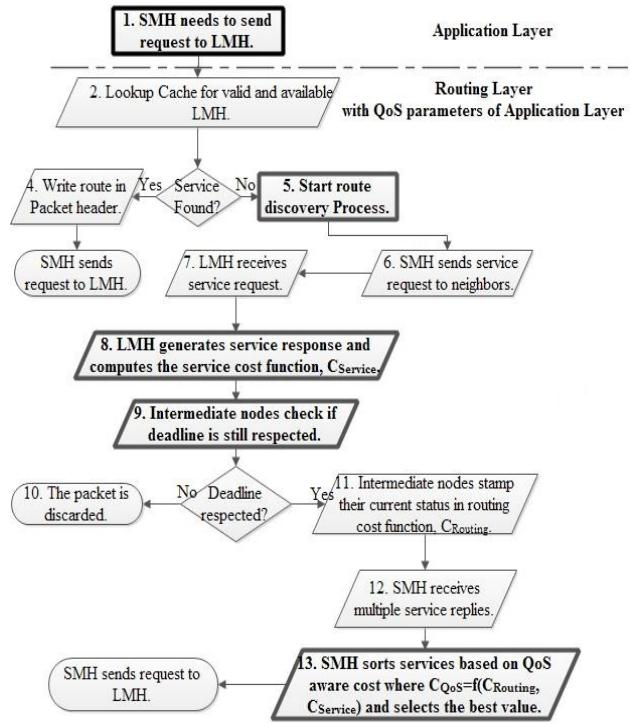

Figure 3. The flow chart of the RTDQS highlighting the cross-layer contribution

At each intermediate node i, the expiration delay is compared to the cost delay contained in the received packet SREP (Figure3, Process9). If the new link cost delay is expired, the packet is discarded. Otherwise, the packet is forwarded to the next intermediate node $i + 1$.

### 4.3. The real-time database service provider functionality

When transmitting a new real-time transaction, the requestor, SMH, sends to a set of service provider nodes, LMH, $k \in \{1,..,N\}$ for an available service, done during the route discovery within multicast address. The RTDQS follows almost the same route discovery method as ED-DSR with different situational modifications. When the RTDBS service provider, LMH, receives a SREQ (Figure3, Process7), it returns back a SREP packet to the SMH via a reversed source route (Figure3, Process8).

In SREP, the $k^{th}$ LMH computes the cost function according to its actual status. The service selection cost function is defined as follows:

$$C_{SS}^k = \alpha C_{Firm}^k + \beta C_{Soft}^k.$$

Where $\alpha$ and $\beta = 1 - \alpha$ are weighting factors of firm and soft cost function, respectively.

The cost function $C_{Firm}$ is for a transaction with a firm real-time constraint. It needs the nearest and the most stable service in order to guarantee the execution of the transaction with a firm real-time constraint. The cost function is defined as follows:

$$C_{Firm}^k = n_{Firm} hP$$

Where $n_{Firm}$ is a normalization factor. $h$ is the number of hops to reach the service provider node. Less number of hops is, shorter distance is. $P$ is the transient overshoot period of the service provider system. P should be shorter in order to reach the steady state after an overloaded period. The overloaded period is characterized by the overshoot of the deadline miss ratio threshold. Less number of hops and shorter transient overshoot period indicate less $C_{Firm}$.

The cost function $C_{Soft}$ is for a transaction with a soft real-time constraint. It requires only a service provider node with less deadline miss ratio and highest remaining energy in order to conserve the nearest service with a rapid stabilization to a transaction with a firm real-time constraint. The cost function is defined as follows:

$$C_{Soft}^k = n_{Soft} \frac{MR}{E_{Residual}}$$

Where $n_{Soft}$ is a normalization factor. $E_{Residual}$ is the remaining energy at the service provider node LMH. MR is the deadline miss ratio of transactions at the LMH. If the deadline miss ratio MR increases and overshoots the threshold value, the transaction will inevitably suffer a longer delay until the system will be stabilized. $C_{Soft}$ increases with more aborted transactions and less remaining energy.

Based on ED-DSR, while a SREP packet is being sent back to the requestor, each intermediate mobile node will stamp its current status in the SREP packet (Figure3, Process11). Finally, at the requestor, the routing agent collects the SREP for a service selection (Figure3, Process12) according to the cost function, described in the next subsection.

### 4.4. The selection of the real-time database service

As in ED-DSR, the requestor waits a certain period of time to collect SREP messages sent from the source data node. Among selected services, the requestor selects one based on a minimum value of $C_{SS}$. However, with a $C_{SS}$ criteria selection, we can have an available service but where intermediate mobile nodes don't have enough resources in order to guarantee the respect of the deadline constraint to transfer the data in time or to maintain the link connection. The cost function should combine the cost computed by intermediate nodes done by the routing protocol ED-DSR and the cost computed by the service provider node, in additive manner. The final QoS cost function should be like this:

$$C_{QoS} = C_{Routing} + C_{SS}$$

The service with the lowest QoS cost $C_{QoS}$ will be chosen (Figure3, Process13).

### 4.5. The multicast routing with the RTDQS

A DSR does not currently support the multicast routing. A multicasting consists of transmitting a packet to a group of mobile nodes identified by a single destination multicast address and hence is intended for a group-oriented computing. The multicast service is employed in areas of a collaborative work e.g. in rescue operations, battlefields video conferencing etc.

The multicasting greatly reduces the transmission cost when sending the same packet to multiple recipients. A multicast packet is typically delivered to all members of its destination group with the same reliability as regular unicast packets. A multicast can reduce communication costs, the link bandwidth consumption, the sender and router processing and the delivery delay. In addition, it can provide a simple and robust communication mechanism when the receiver's individual address is unknown or changeable [8].

International Journal of Database Management Systems ( IJDMS ) Vol.3, No.4, November 2011

Therefore, we have added a new feature to the proposed RTDQS protocol: the multicast routing. When a requestor needs to find an available service provider node in order to process its transaction, he sends a packet to a multicast destination address of several service providers. ED-DSR piggybacks the data from the packet inside a Service Request SREQ targeted at the multicast address. Through an extension of the Route Discovery mechanism, each receiving node then individually examines the destination address of the packet and discards the packet if it is destined to a multicast address to which this node is not subscribed.

## 5. THE SIMULATION MODEL

We have used the Network Simulator NS-2 in our simulations on Linux platform. NS-2 is an object-oriented, event driven simulator. It is suitable for designing new protocols, comparing different protocols or traffic evaluations.

### 5.1. Simulation environment

We simulated a MANET in a 1500m×500m. With a rectangle area, longer distances between the nodes are possible than in a quadratic area, i.e. the packets are sent over more hops. Each node is equipped with an IEEE 802.11 wireless interface in a priority queue of size 50 that drops packets at the queue end in case of overflow.

We defined two groups of mobile nodes according to their resource capacity SMH and LMH. The mobile node groups are defined by the SMH requestor density and LMH service provider density parameters configured for each simulation.

At the beginning of the simulation, SMH nodes start with a starting energy of 50 Joules and LMH with 100 Joules. Since we did not address the problem of consumed energy in idle state, we have only considered energy consumed in transmission and reception modes. As values, we have utilized 1.4 W for a transmission mode and 1 W for a reception mode.

Along the simulation, we vary the state of the LMH in terms of the load and its energy level. We define a set of LMH states in terms of the deadline miss ratio of transactions MR and the transient overshoot period of the service system P in such LMH. LMH are simulated to provide similar service types.

A traffic load between a pair of source-destination (SMH-LMH) is generated by specifying the number of packets per second on the constant bit rate - CBR. Each packet is 512bytes in size.

The mobile nodes move around the simulation area based on the RWP mobility model, with a maximum speed of 2 m/s and a pause time of 10 seconds for SMH, which model a soldier mobility pattern and speeds of up to 20 m/s for LMH, which corresponds more to vehicular movements.

All results reported here are the averages for at least 5 simulation runs. Each simulation runs for 1000 s. During each run, we assume that the node 20 wants to send a real-time traffic to an available service provider node with an expiration delay equals to 15 seconds (firm real-time constraint) and 25 seconds for higher expiration delay (soft real-time constraint). Then, we observe the behaviour of the mobile nodes and the cross-layer service selection proposed protocol performance.
For a service selection, we have chosen the directory-less architecture because we have assumed that our network size is considered small or medium with 20 mobile nodes, relying on [9].

Furthermore, we consider that all LMH provide the same service. The requestor sends to a set of k among N LMH to select the available service.

Simulations of the proposed method were carried out according to a different service provider node density. In this mode, the service transaction load is 5 packets per second, but the service provider density in different simulations ranges from 5 to 40%.

### 5.2. The performance criteria

In order to evaluate the proposed service selection protocol, five important performance metrics are evaluated. They are used to compare the performance of the service selection protocols in the simulation:

**Service availability ratio:** The ratio is defined as a successful contact to this service provider node via the given access information.

**Average service response time:** The average service response is the time elapsed between the instant the service transaction is submitted and the instant the service provider reply is received by the requestor.

**Deadline miss ratio of real-time packet**: The ratio of the real-time data packets that have missed their deadline to be processed by the destination.

**Average energy consumption per bit delivery:** The average is obtained by dividing the sum of the energy consumption of the network by the number of successfully delivered bits.

**Message overhead (packets):** The message overhead is the non-data network level message. It is the average of network layer controlling message packets.

## 6. RESULTS AND DISCUSSIONS

Several simulations are performed using NS-2 network simulator and using parameters described in the last subsection. NS-2 generates a trace file analyzed using a statistical tool developed in AWK. The performance study concerns two versions of the service selection protocol: the closest real-time database service selection protocol which selects the nearest service provider and the proposed RTDQS which refers to our QoS aware service selection for two expiration delays 15s and 25s, which reflect respectively firm and soft real-time constraints.

We study the impact of the service provider density on the service selection protocols performance. This criterion is simulated by varying the number of service providers between {1, 2, 4, 6 and 8} which corresponds to these densities {5%, 10%, 20%, 30% and 40%} with 5 packets per second on the CBR streams.

### 6.1. The service availability ratio

The service availability is defined as a positive service selection reply. It means a successful contact to this service provider via the given access information (i.e. a route to the resolved service provider can be found).

The closest RTD gives almost the same results as our RTDQS with one service provider (5% service provider density). However, at a server density of 40%, the service availability ratio is improved by 64% and 68%, respectively at D=15s (firm real-time constraint) and D= 25s (soft real-time constraint) with our proposed service selection protocol compared to 43% with the closest RTD, as shown in figure 4. In fact, the cost function of our proposed protocol considers the load and the remaining energy of service provider node in the service selection phase.

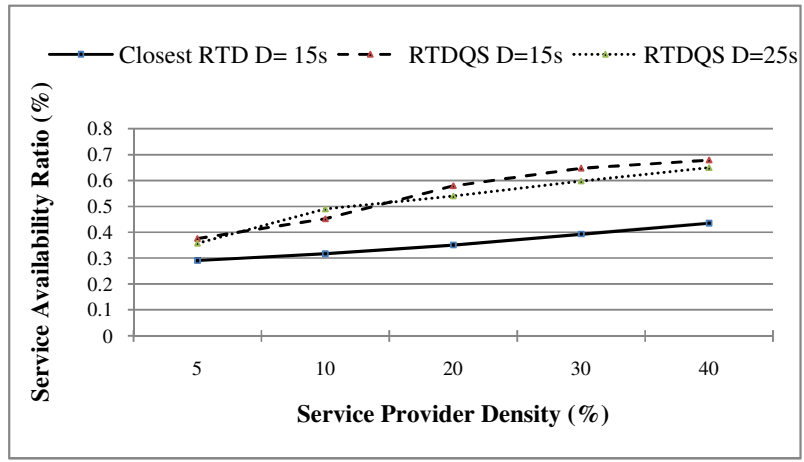

Figure 4.  Service availability ratio of LMH nodes

Our proposed protocol RTDQS selects service provider node depending on the cost function, thereby balancing the load in the network. This load balancing helps to avoid overloaded service providers in the network and reduce the delay for packets, in one hand. Thus for high service provider density (up to 20%), our protocol gives much improved performance than the closest RTD. In the other hand, the RTDQS proposed protocol aims to maintain a balance of energy consumption and workload distribution among service providers in order to prolong their availabilities.

There are few important notes to note between the soft and firm constraint. They follow almost the same attitude. Each transaction has its own function cost adapted to the transaction constraint requirement, as described in the previous subsection.

## 6.2. The service response time

Another commonly used metric is the service response time. It is used to evaluate the service provider performance toward requestor. The average response time of the three approaches with different service density is illustrated in the figure 5. As it can be observed, the performance of hop based approach, the closest RTD, keeps quite steady all through the experiment. The average response time is comparably high compared to our approach because it doesn't consider the system state of the service provider. The service selection of the nearest service provider node is insufficient. It may penalize the transaction response.

The QoS based approach performs the same result as the closest RTD at first, with one service provider. The proposed service selection protocol shows remarkable improvement which proves the efficiency of our approach for both firm and soft real-time constraint with the increase of the service density. When the service provider density increases, more alternatives for transaction processing are featured according to the transaction constraints and the service provider state (load, remaining energy and deadline miss ratio).

In addition, the source node has to select service provider which minimizes the cost function for both the route cost, relative to intermediate nodes, and the service provider cost.

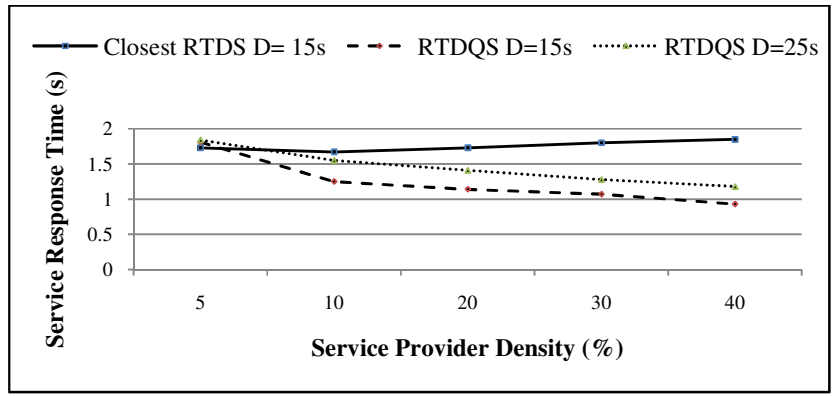

Figure 5. Service response time of LMH nodes

In addition to service state, throughout the route, each intermediate node verifies if the route response packet SREP respects or not the real-time constraint before reaching the source node (the requestor). Thus, the protocol will avoid the network overloading with packets that have expired their deadline in order to reduce energy consumption and thus alleviate the network load and the response time. The figure 5 indicates that the service response time is sensitive to the service provider load and is also dominated by the routing protocol cost, too.

We, also, note that the service response time is a little bit better for lower deadline (D=15s, firm real-time constraint). In fact, the RTDQS selects a service provider which reduces the response delay in order to respect the firm deadline constraint.

### 6.3. The deadline miss ratio packet

We observe and compare the variation of the ratio of packets that have missed their deadlines, while the service provider density is increased.

The figure 6 proves that the closest RTD and the RTDQS provide better performance with high service provider density. However, the RTDQS outperforms the closest RTD for a high service provider density (up to 10%). In fact, with the RTDQS based on the QoS routing protocol ED-DSR, the real-time packets that have expired their deadlines are discarded by the intermediate nodes before reaching the source node (the requestor).

With firm real-time constraint, where D=15s, we note that the packet delivery ratio in time decreases but stills better than the closest RTD. The ratio of the packets sent without the compliance of its real-time constraint is under 10%.

International Journal of Database Management Systems ( IJDMS ) Vol.3, No.4, November 2011

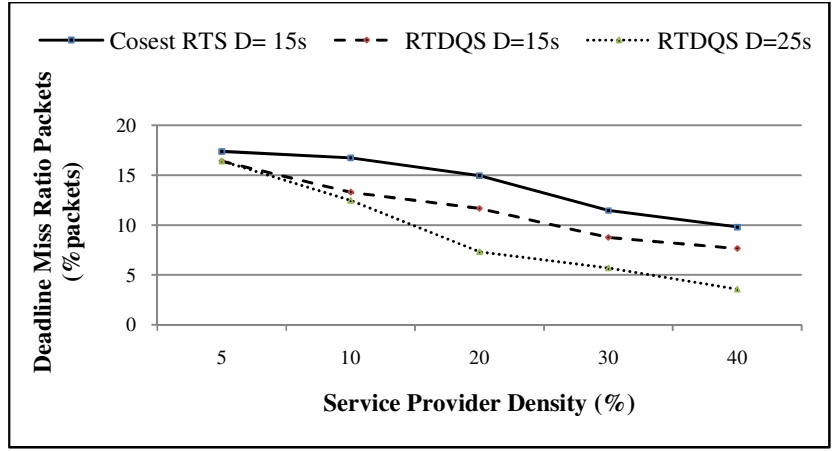

Figure 6. All real-time packets delivery ratio that have missed their real-time constraints

The RTDQS offers best performance for delivering real-time packets in time with a soft real-time constraint, where D=25s. With soft transactions, the first deadline can be neglected in favour of balancing the energy consumption and load over available service providers. The soft transaction is still executed after the first deadline has expired.

### 6.4. The average energy consumption

The figure 7 demonstrates the average energy consumption per bit delivery. It gives an idea about the global energy consumption in the network comparing the RTDQS and the closest RTD under various service densities.

We see that the RTDQS outperforms the closest RTD under different number of service providers, which is mainly due to the benefit of power control in the MAC layer, done by the QoS aware routing protocol ED-DSR.

The RTDQS based on the energy aware ED-DSR routing protocol chooses alternative routes, avoiding the heavily burdened nodes, thus alleviating the explosion in the average energy consumption.

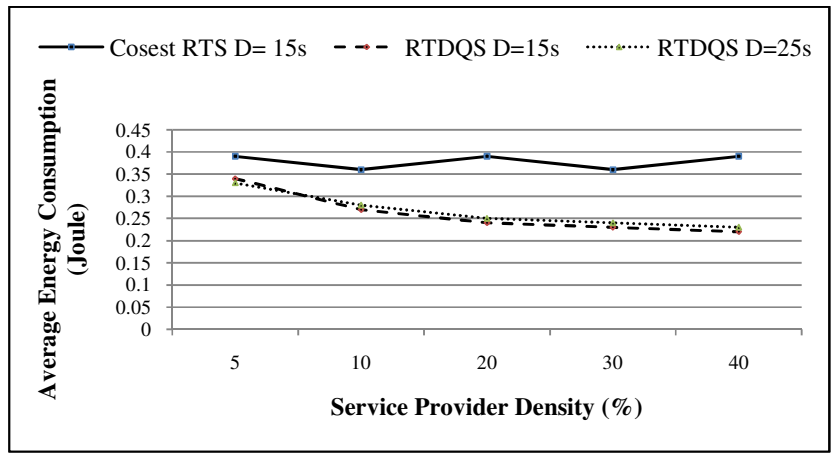

Figure 7. The average energy consumption per bit

113

Changing expiration delay (D=15s for firm real-time constraint and D=25s for soft real-time constraint) for different service provider density had not a significant impact on average energy consumption of the RTDQS, especially for high service density (up to 4 service providers: 20%) because the RTDQS selects path that minimizes the cost function. The energy consumption results of the RTDQS protocol with soft and hard real-time constraints are comparable.

### 6.5. The message overhead

All the non-data network level messages that are transmitted by any node in the network are considered to be message overhead. By measuring overhead as packets, we implicitly count all the routing packets equally.

From the figure 8, we can see that all the service selection protocols follow almost the same attitude with 5% service provider density (which corresponds to one server).

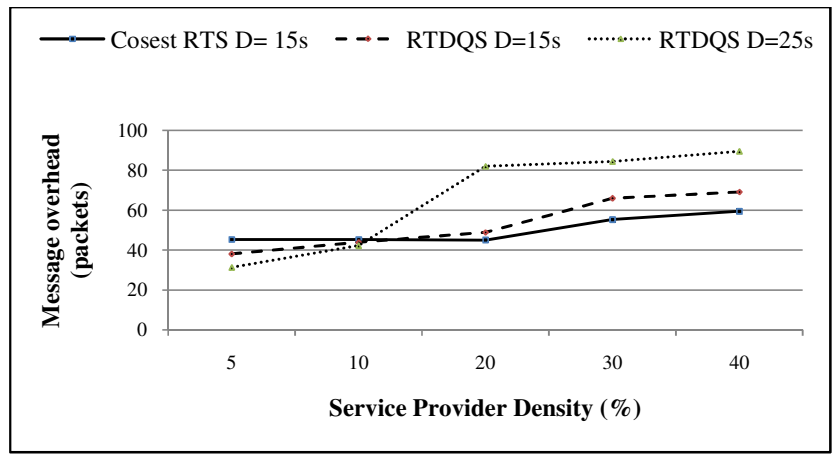

Figure 8. The message overhead

With higher service provider density, the RTDQS with soft real-time constraint generates more overhead messages. In fact, its cost function needs stable service with less deadline miss ratio and with highest residual energy in order to conserve the nearest service with a rapid stabilization to transaction with firm real-time constraint. Therefore, the RTDQS doesn't limit his research to the closest service provider. That's why the RTDQS with firm real-time constraint where D=15s generates less overhead messages than the soft one.

## 7. CONCLUSION AND FUTURE WORK

In this paper, we proposed a cross layer service selection protocol of a real-time database system over MANET. The proposed protocol RTDQS integrates the state information of a RTDBS to those given by the proposed QoS aware routing protocol (an energy and delay-aware DSR routing protocol). Thus, mobile clients may learn about available service providers and routes to them in order to select the suitable one.

Our cross-layer approach helps requestors to select the best possible database service provider. We consider both link state and database service provider state to choose the best and reliable alternative database service provider according to the transaction type. The cost function is defined based on residual energy, load and information state of the database service provider node. The route cost function is defined based on residual energy, queue length, processing and transmission time of intermediate nodes. The database service provider is selected based on minimum value of cost function. The multicast concept is used in order to reduce the transmission cost when sending the same packet to multiple database service provider nodes and thus reduce the number of message overhead.

We compare our approach with the hop based approach: the closest real-time database, the closest RTD. Simulation results indicate that the service selection of the nearest database service provider is insufficient. In fact, respecting the deadline cannot be insured nor guaranteed if the energy resource is exhausted; neither if the real-time database system is overloaded by unpredictable transactions from other requestors.

The closest RTD gives almost the same results as the proposed RTDQS with one database service provider (5% service provider density) for almost the performance criteria. However, for higher database service density (up to 10%), the RTDQS provides more service availability, lower service response time, achieves lower energy dissipation per bit of data delivery and lower deadline miss ratio.

In the future, we plan to study alternative commit protocols and then incorporate the concurrency control into our approach. A suitable CC algorithm for MANET databases should be energy-efficient in order to improve high concurrency and avoid wasting limited system resources. Otherwise, we plan to evaluate the proposed protocol with other simulation mode such as varying service transaction load or network density.

**AUTHORS**

**Jihen DRIRA REKIK** is a Ph. D. student working on QoS guarantee of real-time database system over mobile Ad-hoc Networks at the National School of Computer Sciences of Tunis. She received her Master in Communication systems from National Engineering School of Tunis (ENIT) and her engineering degree in computer network from National Institute of Applied Science and Technology (INSAT). Now, as a Ph. D. candidate, her research interests include energy efficiency, delay guarantee, mobile ad-hoc networks and real-time database system.

**Leïla BACCOUCHE** received her Ph. D. in computer science from the National Polytechnic Institute of Grenoble in France in 1996. She is an assistant professor at the National Institute of Applied Science and Technology in Tunisia in the computer science and mathematics department. Her research interest is related to real-time systems and real-time databases and includes scheduling, quality of service, feedback control and mobile wireless and ad-hoc networks.

**Henda BEN GHEZALA** is currently a Professor of Computer Science in the department of Informatics at the National School of Computer Sciences of Tunis. She leads a Master degree in 'ICIS'. She is the president of University of Manouba. Her research interests lie in the areas of information modeling, databases, temporal data modeling, object-oriented analysis and design, requirements engineering and specially change engineering, method engineering. She is the director of the RIADI laboratory.